\documentclass[aps,floatfix,showpacs]{revtex4}
\usepackage{graphicx,amsfonts,amssymb,amsbsy}
\usepackage{pstricks}
\usepackage{graphicx}
\usepackage{slashed}
\usepackage{bm}

\usepackage{amssymb}

\begin{document}

\title{\bf Deconfinement transition in protoneutron stars: analysis within the Nambu-Jona-Lasinio model}

\author{G. Lugones$^1$, T. A. S. do Carmo$^1$, A. G. Grunfeld$^{2,3,4}$, N. N. Scoccola$^{2,3,5}$ }

 \affiliation{ $^1$ Universidade Federal do ABC, Centro de Ciencias
 Naturais e Humanas, Rua Santa Ad\'elia, 166, 09210-170, Santo Andr\'e, Brazil\\
 $^2$ CONICET, Rivadavia 1917, (1033) Buenos Aires, Argentina.\\
 $^3$ Departmento de F\'\i sica, Comisi\'on Nacional de
 Energ\'{\i}a At\'omica, (1429) Buenos Aires, Argentina.\\
  $^4$ Department of Physics, Sultan Qaboos University, P.O.Box: 36 Al-Khode 123 Muscat, Sultanate of Oman \\
 $^5$ Universidad Favaloro, Sol{\'\i}s 453, (1078) Buenos Aires, Argentina.}

\vskip5mm
\begin{abstract}
We study the effect of color superconductivity and neutrino trapping on the deconfinement transition
of hadronic matter into quark matter in a protoneutron star.  To describe
the strongly interacting matter a two-phase picture is adopted. For the hadronic phase
we use different parameterizations of a non-linear Walecka model
which includes the whole baryon octet. For the quark matter phase we use an $SU(3)_f$
Nambu-Jona-Lasinio effective model which includes color superconductivity. We impose color
and flavor conservation during the transition in such a way that just deconfined
quark matter is transitorily out of equilibrium with respect to weak interactions.
We find that deconfinement is more difficult for small neutrino content and it is easier
for lower temperatures although these effects are
not too large. In addition they will tend to cancel each other as the protoneutron star cools
and deleptonizes, resulting a transition density that is roughly constant along the evolution
of the protoneutron star.  According to these results the deconfinement transition is favored after substantial cooling and contraction of the protoneutron star.
\end{abstract}


\maketitle

\section{Introduction}

It is currently a matter of speculation the actual occurrence of
quark matter during protoneutron star (PNS) evolution. The
standard  scenario for  the birth of neutron stars indicates that
these objects are formed as consequence of the  gravitational
collapse and supernova explosion of a massive  star
\cite{Burrows1986,Keil1995,Pons1999}. Initially, PNSs are very hot
and lepton-rich objects, where neutrinos are temporarily trapped.
During the first tens of seconds of evolution the PNS evolves to
form a cold \mbox{($T<10^{10}$ K)} catalyzed neutron star
\cite{Burrows1986,Keil1995,Pons1999}. As neutrinos are  radiated,
the  lepton -  per -  baryon content  of matter  goes down  and
the neutrino chemical potential tends to essentially zero in $\sim
50$ seconds \cite{Pons1999}. Deleptonization is fundamental for
quark  matter  formation  inside  neutron stars, since it has been
shown  that  the  presence  of trapped neutrinos in hadronic
matter strongly disfavors the deconfinement transition
\cite{Lugones1998,Lugones1999}. In fact, neutrino trapping makes
the density for the deconfinement transition to be higher than in
the case of neutrino-free hadronic matter. As a consequence, the
transition could be delayed  several seconds  after the bounce of
the stellar core. However, the calculations presented in
\cite{Lugones1998,Lugones1999} were performed employing the MIT
Bag model for the description of quark matter and did not include
the effect of color superconductivity. As we shall see in the
present work, the use of the Nambu-Jona-Lasinio model and the
inclusion of color superconductivity may change qualitatively the
effect of neutrino trapping in the deconfinement conditions.

As emphasized in earlier works
\cite{Madsen1994,IidaSato1998,Lugones1998,Lugones1999,Bombaci2004,Lugones2005,Bombaci2007,Bombaci2009},
an important characteristic of the deconfinement transition in
neutron stars, is that just deconfined quark matter is
transitorily out of equilibrium with respect to weak interactions.
In fact, depending on the temperature, the transition should begin
with the quantum or thermal nucleation of a small quark-matter
drop near the center of the star. On the other hand, the flavor
composition of hadronic matter in  $\beta$-equilibrium is
different from that of a $\beta$-stable quark-matter drop. Roughly
speaking, the direct formation of a $\beta$-stable quark-drop with
$N$ quarks will need the almost simultaneous conversion of $\sim
N/3$ {up and down} quarks into strange quarks, a process which is
strongly suppressed with respect to the formation of a non
$\beta$-stable drop by a factor $\sim G_{\mathrm{Fermi}}^{2N/ 3}$.
For typical values of the critical-size $\beta$-stable drop ($N
\sim 100-1000$ \cite{IidaSato1998}) the suppression factor is
actually tiny. Thus, quark flavor must be conserved during the
deconfinement transition
\cite{Madsen1994,IidaSato1998,Lugones1998,Lugones1999,Bombaci2004,Lugones2005,Bombaci2007}.
When color superconductivity is included together with flavor
conservation, the most likely configuration of the just deconfined
phase is 2SC provided the pairing gap is large enough
\cite{Lugones2005}. The relevance of this 2SC intermediate phase
(a kind of activation barrier) has been analyzed for deleptonized
neutron stars \cite{Bombaci2007,Lugones2009} but not for  hot and
lepton-rich objects like PNSs.

In the present paper we shall analyze the deconfinement transition
in protoneutron star conditions employing the Nambu-Jona-Lasinio
model in the description of quark matter. For the hadronic phase
we shall use a model based on a relativistic Lagrangian of hadrons
interacting via the exchange of $\sigma$, $\rho$, and $\omega$
mesons \cite{GM1}. For simplicity, the analysis will be made in
bulk, i.e. without taking into account the energy cost due to
finite size effects in creating a drop of deconfined quark matter
in the hadronic environment.

The present article is organized as follows. In Sec. II we
present the main aspects of the non--linear Walecka model
describing the hadronic phase. In Sec. III we present the
generalities of the model we use for the quark phase. In Sec. IV
we show our numerical results and finally in Sec. V we discuss
our results and present the conclusions.

%------------------------------------
\section{The hadronic phase}
%------------------------------------
\label{hphase}

For the hadronic phase we shall use a non-linear Walecka model
(NLWM) \cite{qhd,GM1,Menezes} which includes the whole baryon
octet, electrons and electron neutrinos in equilibrium under weak
interactions. The Lagrangian of the model is given by
\begin{equation}
{\cal L}={\cal L}_{B}+{\cal L}_{M}+{\cal L}_{L},
\label{octetlag}
\end{equation}
where the indices $B$, $M$ and $L$ refer to baryons, mesons and
leptons respectively. For the baryons we have
\begin{eqnarray}
{\cal L}_B= \sum_B \bar \psi_B
\bigg[\gamma^\mu\left (i\partial_\mu -
g_{\omega B} \ \omega_\mu- g_{\rho B} \ \vec \tau \cdot \vec \rho_\mu \right)
%\nonumber\\
-(m_B-g_{\sigma B} \ \sigma)\bigg]\psi_B,
\end{eqnarray}
with $B = n$, $p$, $\Lambda$, $\Sigma^{+}$, $\Sigma^{0}$, $\Sigma^{-}$,
$\Xi^{-}$, and $\Xi^{0}$. The contribution of the mesons $\sigma$,
$\omega$ and $\rho$ is given by
\begin{eqnarray}
{\cal L}_{M} &=& \frac{1}{2} (\partial_{\mu} \sigma \ \! \partial^{\mu}\sigma -m_\sigma^2 \ \! \sigma^2)
- \frac{b}{3} \ \! m_N\ \! (g_\sigma\sigma)^3 -\frac{c}{4} \ (g_\sigma \sigma)^4
\nonumber\\
& &
-\frac{1}{4}\ \omega_{\mu\nu}\ \omega^{\mu\nu} +\frac{1}{2}\ m_\omega^2 \ \omega_{\mu}\ \omega^{\mu}
%\nonumber \\ & &
-\frac{1}{4}\ \vec \rho_{\mu\nu} \cdot \vec \rho\ \! ^{\mu\nu}+
\frac{1}{2}\ m_\rho^2\  \vec \rho_\mu \cdot \vec \rho\ \! ^\mu,
\end{eqnarray}
where the coupling constants are $g_{\sigma B}=x_{\sigma B}~
g_\sigma$, $g_{\omega B}=x_{\omega B}~ g_{\omega}$ and $g_{\rho
B}=x_{\rho B}~ g_{\rho}$. The ratios $x_{\sigma B}$, $x_{\omega
B}$ and $x_{\rho B}$ are equal to $1$ for the nucleons and acquire
different values for the other baryons depending on the
parametrization (see Table I). The leptonic sector is included as
a free Fermi gas of electrons and electron neutrinos in chemical
equilibrium with the other particles.

\begin{table*}[t!]
\centering
\begin{tabular}{l|c|c|c|c|c|c|c|c|cc}
\hline\hline
Label & composition & $x_{\sigma} = x_{\rho}$ & $x_{\omega}$ & $(g_{\sigma}/m_{\sigma})^2 $
& $(g_{\omega}/m_{\omega})^2$     &  $(g_{\rho}/m_{\rho})^2 $  &  $b$   &   $c$  &  $M_{max}$ \\
&  &  &  & $~\mathrm{[fm^2]}$ & $ ~\mathrm{[fm^2]}$     &  $ ~\mathrm{[fm^2]}$  &   &    &  \\
\hline
GM 1 & baryon octet + $e^-$  & 0.6 & 0.653    &  11.79  & 7.149 & 4.411   & 0.002947  & -0.001070 & 1.78 $M_{\odot}$ \\
% GM 3 & baryon octet + $e^-$ & .. &     &  9.927  & 4.820 & 4.791   & %0.008659  & -0.002421 & 1.53 $M_{\odot}$ \\
GM 4 & baryon octet + $e^-$ & 0.9 & 0.9    &  11.79  & 7.149 & 4.411   & 0.002947  & -0.001070 & 2.2 $M_{\odot}$ \\
\hline\hline
\end{tabular}
\caption{Parameters of the hadronic equation of state. For each parametrization we give the maximum mass $M_{max}$
of a hadronic star.}
\label{setshadronicos}
\end{table*}

There are five constants in the model that are determined by the
properties of nuclear matter, three that determine the nucleon
couplings to the scalar, vector and vector-isovector mesons
$g_{\sigma}/m_{\sigma}$, $g_{\omega}/m_{\omega}$,
$g_{\rho}/m_{\rho}$, and two that determine the scalar self
interactions $b$ and $c$. It is assumed that all hyperons in the
octet have the same coupling than the $\Lambda$. These couplings
are expressed as a ratio to the nucleon couplings mentioned above,
that we thus simply denote $x_\sigma$, $x_\omega$ and $x_\rho$. In
the present work we use two parameterizations for the constants.
One of them is the standard parameterization GM1 given by
Glendenning--Moszkowski \cite{GM1}, as shown in Table I. This
parametrization employs ``low'' values for $x_\sigma$,  $x_\omega$
and  $x_\rho$. The parametrization GM4 employs larger values of these couplings. This makes the EOS
stiffer and increases the maximum mass of hadronic stars to 2.2 $M_{\odot}$, see Table I.

The derivation of the equations describing the model is given in detail in \cite{mfm}.
The total pressure $P$  and mass - energy  density $\rho$ are given by:
\begin{eqnarray}
P =  \sum_{i=B,L}{ P_i }
+ {1\over{2}} \bigg({g_{\omega}\over{m_{\omega}}} \bigg)^2 \rho_{B}^{'2}
- {1\over{2}} \bigg( {g_{\sigma}\over{m_{\sigma}}} \bigg)^{-2}
( g_{\sigma} \sigma )^2
- {1 \over{3}} b m_n (g_{\sigma} \sigma)^3
- {1 \over{4}} c (g_{\sigma} \sigma)^4
+ {1\over{2}}  \bigg({g_{\rho}\over{m_{\rho}}} \bigg)^2
\rho_{I_3}^{'2} ,
\end{eqnarray}
\begin{eqnarray}
\rho =  \sum_{i=B,L}{ \rho_i }
+ {1\over{2}} \bigg({g_{\omega}\over{m_{\omega}}} \bigg)^2 \rho_{B}^{'2}
+ {1\over{2}} \bigg( {g_{\sigma}\over{m_{\sigma}}} \bigg)^{-2}
(g_{\sigma} \sigma)^2
+ {1 \over{3}} b m_n (g_{\sigma} \sigma)^3
+ {1 \over{4}} c (g_{\sigma} \sigma)^4
+ {1\over{2}}  \bigg( {g_{\rho}\over{m_{\rho}}} \bigg)^2
\rho_{I_3}^{'2} .
\end{eqnarray}
Here $P_i$  and $\rho_i$  are the  expressions for  a Fermi gas of relativistic, non-interacting particles:
\begin{equation}
P_i =  {1 \over{3}} {g_{i} \over{(2 \pi)^3} }
\int { d^{3}p
\;    { {p^2} \over{(p^2 + m_i^{*2})^{1/2}} }
\;     ( f_i(T) +  \bar{f}_i(T) )} ,
\label{Pres}
\end{equation}
\begin{equation}
\rho_i =  {g_{i} \over{(2 \pi)^3} }
\int { d^{3}p
\;    {(p^2 + m_i^{*2})^{1/2}}
\;    ( f_i(T) +  \bar{f}_i(T) )} ,
\label{Ener}
\end{equation}

where  $f_i(T)$  and  $\bar{f}_i(T)$   are  the  Fermi  -   Dirac
distribution   functions   for   particles   and    antiparticles
respectively:
\begin{equation}
f_i(T) =
(exp ( [ (p^2 + m_i^{*2})^{1/2} - \mu_i^* ] / T )  + 1 )^{-1} ,
\end{equation}
\begin{equation}
\bar{f}_i(T) =
(exp ( [ (p^2 + m_i^{*2})^{1/2} + \mu_i^* ] / T )  + 1 )^{-1} .
\end{equation}
Note that for  baryons we use, instead of  masses $m_i$ and chemical potentials $\mu_i$,  ``effective'' masses $m_i^{*}$  and
chemical potentials $\mu_i^*$ given by:
\begin{equation}
m_i^{*} = m_i + x_{\sigma i} (g_{\sigma} \sigma) ,
\end{equation}
\begin{equation}
\mu_i^* = \mu_i
- x_{\omega i} \bigg({g_{\omega}\over{m_{\omega}}} \bigg)^2 \rho^{'}_{B}
- x_{\rho i} I_{3 i} \bigg( {g_{\rho}\over{m_{\rho}}} \bigg)^2
\rho^{'}_{I_3} ,
\end{equation}
where $I_{3 i}$ is the third component of the  isospin of each baryon.

The weighted  isospin density  $\rho^{'}_{I_3}$ and  the weighted
baryon density $\rho^{'}_{B}$ are given by:
\begin{equation}
\rho^{'}_{I_3}  = \sum_{i=B}{ x_{\rho i} I_{3 i} n_i} ,
\end{equation}
\begin{equation}
\rho^{'}_{B}  = \sum_{i=B}{ x_{\omega i} n_i} ,
\end{equation}
being $n_i$ the particle number density of each baryon:
\begin{equation}
n_i =   {g_{i} \over{(2 \pi)^3} }
\int { d^{3}p   \;   ( f_i(T) - \bar{f}_i(T) )} .
\label{num}
\end{equation}

The mean field $g_{\sigma} \sigma$ satisfies the equation:
\begin{equation}
\bigg({g_{\sigma}\over{m_{\sigma}}} \bigg)^{-2}
(g_{\sigma} \sigma)
+  b m_n (g_{\sigma} \sigma)^2
+  c (g_{\sigma} \sigma)^3
=  \sum_{i=B}{ x_{\sigma i} n^s_i} ,
\end{equation}
where $n^s_i$ is the scalar density:
\begin{equation}
n^s_i =
{g_{i} \over{(2 \pi)^3} }
\int { d^{3}p
\;    { {m_i^{*}} \over{(p^2 + m_i^{*2})^{1/2}} }
\;     ( f_i(T) +  \bar{f}_i(T) )
} .
\end{equation}

The hadron  phase is  assumed to  be charge  neutral and  in chemical equilibrium under weak interactions.
Electric charge neutrality states:
\begin{equation}
n_p + n_{\Sigma^{+}} -  n_{\Sigma^{-}} - n_{\Xi^{-}}  - n_{e} = 0 .
\end{equation}
Chemical weak  equilibrium in  the presence  of trapped  electron
neutrinos implies that the chemical potential $\mu_i$  of each
baryon in the hadron phase is given by:
\begin{equation}
\mu_i =  q_B \mu_n - q_e (\mu_e - \mu_{\nu_e} ) ,
\end{equation}
where $q_B$  is its  baryon charge  and $q_e$  is its  electric
charge.    For  simplicity  we are assuming that  muon and tau
neutrinos are not present in the system, and their chemical potentials are  set
to zero.

All the above equations can be solved numerically by  specifying
three thermodynamic quantities, e.g. the temperature $T$, the mass-energy density $\rho$
and the chemical potential of electron neutrinos in the hadronic phase $\mu_{\nu_e}^H$.

%--------------------------------
\section{The quark matter phase}
%--------------------------------
\label{qphase}

In order to study the just deconfined quark matter phase we use an
$SU(3)_f$ NJL effective model which also includes color
superconducting quark-quark interactions. The corresponding
Lagrangian is given by
\begin{eqnarray}
{\cal L} &=& \bar \psi \left(i \rlap/\partial - \hat m \right)
\psi
%\nonumber \\ & &
+ G \sum_{a=0}^8 \left[ \left( \bar \psi \ \tau_a \ \psi \right)^2
+ \left( \bar \psi \ i \gamma_5 \tau_a \ \psi \right)^2 \right]
%\\
% & & \qquad
+ 2H \!\!
\sum_{A,A'=2,5,7} \left[ \left( \bar \psi \ i \gamma_5
\tau_A \lambda_{A'} \ \psi_C \right) \left( \bar \psi_C \ i
\gamma_5 \tau_A \lambda_{A'} \ \psi \right) \right]
%\nonumber
\label{action}
\end{eqnarray}
where $\hat m=\mathrm{diag}(m_u,m_d,m_s)$ is the current mass
matrix in flavor space. In what follows we will work in the
isospin symmetric limit $m_u=m_d=m$. Moreover, $\tau_i$ and
$\lambda_i$ with $i=1,..,8$ are the Gell-Mann matrices
corresponding to the flavor and color groups respectively, and
$\tau_0 = \sqrt{2/3}\ 1_f$. Finally, the charge conjugate spinors
are defined as follows: $\psi_C = C \ \bar \psi^T$ and $\bar
\psi_C = \psi^T C$, where $\bar \psi = \psi^\dagger \gamma^0$ is
the Dirac conjugate spinor and $C=i\gamma^2 \gamma^0$.

To be able to determine the relevant thermodynamical quantities we
have to obtain the grand canonical thermodynamical potential at
finite temperature $T$ and chemical potentials $\mu_{fc}$. Here,
$f=(u,d,s)$ and $c=(r,g,b)$ denotes flavor and color indices
respectively. For this purpose, starting from Eq. (\ref{action}),
we perform the usual bosonization of the theory. This can be done
by introducing scalar and pseudoscalar meson fields $\sigma_a$ and
$\pi_a$ respectively, together with the bosonic diquark field
$\Delta_A$. In this work we consider the quantities obtained
within the mean field approximation (MFA). Thus, we only keep the
non-vanishing vacuum expectation values of these fields and drop
the corresponding fluctuations. For the meson fields this implies
$\hat \sigma = \sigma_a \tau_a =
\textrm{diag}(\sigma_u,\sigma_d,\sigma_s)$ and $\pi_a=0$.
Concerning the diquark mean field, we will assume that in the
density region of interest only the 2SC phase might be relevant.
Thus, we adopt the ansatz $\Delta_5 = \Delta_7 = 0$, $\Delta_2 =
\Delta$. Integrating out the quark fields and working in
the framework of the Matsubara and Nambu-Gorkov formalism we
obtain the following MFA quark thermodynamical potential
(a detailed procedure of calculation can be found in Refs. \cite{
Huang:2002zd,Ruester:2005jc,Blaschke:2005uj} )
\begin{eqnarray}
\Omega^{MFA}_q(T,
\mu_{fc},\sigma_u,\sigma_d,\sigma_s,|\Delta|) =
\frac{1}{\pi^2}\int_0^\Lambda dk \; k^2 \sum_{i=1}^9
\omega(x_i,y_i) +
% \nonumber\\
\frac{1}{4G}(\sigma_u^2+\sigma_d^2+\sigma_s^2)  +
\frac{|\Delta|^2}{2H},
\end{eqnarray}
where $\Lambda$ is the cut-off of the model and $\omega(x,y)$ is
defined by
\begin{eqnarray}
\omega(x,y) = - \left[ x + T\ln[1+e^{-(x-y)/T}]  + T\ln[1+e^{-(x+y)/T}] \right] \ ,
\end{eqnarray}
with
\begin{eqnarray}
x_{1,2} = E \ \ , \ \ x_{3,4,5} = E_s \ \ , \ \ x_{6,7} =
\sqrt{\bigg[ E +  \frac{(\mu_{ur} \pm \mu_{dg})}{2} \bigg]^2 +
\Delta^2} \ \ , \ \ x_{8,9} = \sqrt{\bigg[ E +  \frac{(\mu_{ug}
\pm \mu_{dr})}{2} \bigg]^2 + \Delta^2} \ , \nonumber
\end{eqnarray}
\begin{eqnarray}
y_1  = \mu_{ub} \ , \ \ y_2 = \mu_{db} \ , \ \ y_{3} = \mu_{sr} \
, \ \ y_{4} = \mu_{sg} \ , \ \ y_{5} = \mu_{sb} \ , \ \ y_{6,7} =
\frac{(\mu_{ur}-\mu_{dg})}{2} \ , \ \ y_{8,9} = \frac{\mu_{ug} -
\mu_{dr}}{2} \ .
\end{eqnarray}
Here, $E = \sqrt{ k^2+ M^2}$ and $E_s=\sqrt{  k ^2 +
M_s^2}$, where $M_f = m_f + \sigma_f$. Note that in the isospin
limit we are working $\sigma_u = \sigma_d = \sigma$ and, thus, $M_u = M_d = M$.

The total thermodynamical potential of the quark matter phase (QMP) is obtained
by adding to $\Omega_{MFA}$  the contribution of the leptons. Namely,
\begin{equation}
\Omega_{QMP}(T,
\mu_{fc},\mu_e,\mu_{\nu_e},\sigma,\sigma_s,|\Delta|) =
\Omega^{MFA}_q(T, \mu_{fc},\sigma,\sigma_s,|\Delta|) +
\Omega_e(T,\mu_e)+ \Omega_{\nu_e}(T,\mu_{\nu_e}) -
\Omega_\textrm{\tiny vac} \label{QMP}
\end{equation}
where $\Omega_e$ and $\Omega_{\nu_e}$ are the thermodynamical potentials
of the electrons and neutrinos, respectively. For them we use the expression corresponding to a free gas of ultra-relativistic fermions
\begin{equation}
\Omega_l(T,\mu_{l}) = - \gamma_l \left( \frac{\mu_l^4}{24 \pi^2} + \frac{\mu_l^2T^2}{12} + \frac{7\pi^2T^4}{360} \right),
\nonumber
\end{equation}
where $l = e, \nu_e$ and the degeneracy factor is $\gamma_e = 2$ for electrons and
$\gamma_{\nu_e} = 1$ for neutrinos. Notice that in Eq.(\ref{QMP}) we have subtracted the constant
$\Omega_\textrm{\tiny vac}$ in order to have a vanishing pressure
at vanishing temperature and chemical potentials.

From the grand thermodynamic potential $\Omega_{QMP}$ we can readily obtain  the pressure $P = - \Omega_{QMP}$, the
number density of quarks of each flavor and color $n_{fc} =  - { \partial \Omega_{QMP} }/{\partial \mu_{fc} }$,
the number density of electrons  $n_{e} = - {\partial \Omega_{QMP} }/{\partial \mu_e}$, and the number density
of electron neutrinos  $n_{\nu_e} = - { \partial \Omega_{QMP} }/{\partial \mu_{\nu_e}}$. The corresponding
number densities of each flavor, $n_f$, and of each color, $n_c$, in the quark phase are given
by $n_f = \sum_{c} n_{fc}$  and  $n_c = \sum_{f} n_{fc}$ respectively.  The baryon number
density reads $n_B = \frac{1}{3} \sum_{fc} n_{fc} = (n_u + n_d + n_s)/3$. Finally, the Gibbs free energy per baryon is
\begin{equation}
g_\textrm{\scriptsize quark}= \frac{1}{n_B}\left(\sum_{fc} \mu_{fc} \ n_{fc} + \mu_e \ n_e +  \mu_{\nu_e} \ n_{\nu_e}
\right).
\label{g_quark}
\end{equation}

For the NJL model we use two sets of constants shown in Table
\ref{sets}. The sets 1 and 2 were taken from \cite{Rehberg:1995kh}
and \cite{Hatsuda:1994pi} respectively, but without the 't Hooft
flavor mixing interaction. The procedure, obtained from
\cite{Buballa2005} is to keep $\Lambda$ and $m$ fixed, then tune
the remaining parameters $G$ and $m_s$ in order to reproduce $M
=367.6$ MeV and $M_s=549.5$ MeV at zero temperature and density.
The resulting parameter sets are given in Table \ref{sets}.

%%%%%%%%%%%%%%%%%%%%%%%%%%%%%%%%%%%%%%%%%%%%%%%%%%%%%%%%
\begin{table*}[t!]
%\begin{table*}[h]
\centering
\begin{tabular}{c|ccccc}
\hline\hline & $m_{u,d} $ [Mev] & $m_s$ [Mev] & $\Lambda$ [Mev] &
$G\Lambda^2$&
$H/G$\\
\hline
set 1 & 5.5 & 112.0  & 602.3 & 4.638 & 3/4 \\
set 2 & 5.5 & 110.05 & 631.4 & 4.370 & 3/4 \\
\hline
\end{tabular}
\caption{The two sets of NJL parameters.} \label{sets}
\end{table*}

%%%%%%%%%%%%%%%%%%%%%%%%%%%%%%%%%%%%%%%%%%%%%%%%%%%%%%%%%%

In order to derive a quark matter EOS from the above formulae it
is necessary to impose a suitable number of conditions on the
variables  $\{\mu_{fc}\}, \mu_e, \mu_{\nu_e},\sigma, \sigma_s$ and $\Delta$.
Three of these conditions are consequences from the fact that the
thermodynamically consistent solutions correspond to the
stationary points of $\Omega$ with respect to $\sigma$,
$\sigma_s$, and $\Delta$. Thus, we have
\begin{eqnarray}
\partial\Omega_{QMP}/\partial\sigma =0
\qquad , \qquad
\partial\Omega_{QMP}/\partial\sigma_s =0
\qquad , \qquad
\partial\Omega_{QMP}/\partial|\Delta|=0.
\label{gapeq}
\end{eqnarray}

To obtain the remaining conditions one must specify the physical
situation in which one is interested in. As in previous works
\cite{Madsen1994,IidaSato1998,Lugones1998,Lugones1999,Bombaci2004,Lugones2005,Bombaci2007},
we are dealing here with just deconfined quark matter that is
temporarily out of chemical equilibrium under weak interactions.
The appropriate condition in this case is flavor conservation
between hadronic and deconfined quark matter. This can be written
as
\begin{equation}
Y^H_f = Y^Q_f   \;\;\;\;\;\; f=u,d,s,e, \nu_e \label{flavor}
\end{equation}
being $Y^H_f \equiv n^H_f / n^H_B$ and  $Y^Q_i \equiv n^Q_f /
n^Q_B$ the abundances of each particle in the hadron and quark
phase respectively. In other words, the just deconfined quark
phase must have the same ``flavor'' composition than the
$\beta$-stable hadronic phase from which it has been originated.
Notice that, since the hadronic phase is assumed to be
electrically neutral, flavor conservation ensures automatically
the charge neutrality of the just deconfined quark phase.  The
conditions given in Eq. (\ref{flavor}) can be combined to obtain
\begin{equation}
n_d = \xi ~ n_u
\qquad , \qquad
n_s = \eta ~ n_u
\qquad , \qquad
n_{\nu_e} = \kappa ~ n_u
\qquad , \qquad
3 n_{e} = 2 n_{u} - n_{d} - n_{s} ,\label{h3}
\end{equation}
\noindent where $n_i$ is the particle number density of the
$i$-species in the quark phase. The quantities $\xi \equiv Y^H_d /
Y^H_u$, $\eta \equiv Y^H_s / Y^H_u$  and $\kappa \equiv Y^H_{\nu_e} / Y^H_u$ are functions of the
pressure and temperature, and they characterize the composition of
the hadronic phase. These expressions are valid for \textit{any}
hadronic EOS. For hadronic matter containing $n$, $p$, $\Lambda$,
$\Sigma^{+}$, $\Sigma^{0}$, $\Sigma^{-}$, $\Xi^{-}$ and $\Xi^{0}$,
we have
\begin{eqnarray}
\xi &=& \frac{n_p  +  2  n_n  + n_{\Lambda} + n_{\Sigma^{0}} +  2
n_{\Sigma^{-}}  + n_{\Xi^{-}}}{2  n_p  +  n_n  +  n_{\Lambda} + 2
n_{\Sigma^{+}} + n_{\Sigma^{0}}  +  n_{\Xi^{0}}}, \label{xi} \\
\eta &=& \frac{n_{\Lambda}  + n_{\Sigma^{+}} + n_{\Sigma^{0}}  +
n_{\Sigma^{-}} + 2 n_{\Xi^{0}} + 2 n_{\Xi^{-}}}{2  n_p  +  n_n  +
n_{\Lambda} + 2 n_{\Sigma^{+}} + n_{\Sigma^{0}}  +  n_{\Xi^{0}}}, \\
\kappa &=& \frac{n^H_{\nu_e}}{2  n_p  +  n_n  +
n_{\Lambda} + 2 n_{\Sigma^{+}} + n_{\Sigma^{0}}  +  n_{\Xi^{0}}}.
\label{eta}
\end{eqnarray}
Additionally, the deconfined phase must be locally colorless; thus
it must be composed by an equal number of red, green and blue
quarks
\begin{equation}
n_r = n_g = n_b
 \label{colorless}.
\end{equation}
Also, $ur$, $ug$, $dr$, and $dg$ pairing will happen provided that
$|\Delta|$ is nonzero, leading to
\begin{equation}
n_{ur}=n_{dg}
\qquad , \qquad
n_{ug}=n_{dr}.
\label{pairing}
\end{equation}

In order to have all Fermi levels at the same value, we consider
\cite{Lugones2005}
\begin{eqnarray}
n_{ug} = n_{ur}
\qquad , \qquad
n_{sb} = n_{sr}.
\label{equalfermilevels}
\end{eqnarray}
These two equations, together with Eqs. (\ref{colorless}) and
(\ref{pairing}) imply that $n_{ur}=n_{ug}=n_{dr}=n_{dg}$ and
$n_{sr}=n_{sg}=n_{sb}$ \cite{Lugones2005}.

Finally, including the conditions Eqs.(\ref{gapeq}) we have
13 equations involving the 14 unknowns ($\sigma$,
$\sigma_s$, $|\Delta|$, $\mu_e$, $\mu_{\nu_e}$ and
$\{\mu_{fc}\}$). For given value of one of the chemical potentials
(e.g. $\mu_{ur}$), the set of equations can be solved once the
values of the parameters  $\xi$, $\eta$, $\kappa$  and the
temperature $T$ are given. Instead of $\mu_{ur}$, we can provide a
value of the Gibbs free energy per baryon $g_\textrm{\scriptsize
quark}$ or the pressure $P$ and solve simultaneously Eqs.
(\ref{h3})-(\ref{equalfermilevels}) together with Eq.
(\ref{gapeq}) in order to obtain $\sigma$, $\sigma_s$, $|\Delta|$,
$\mu_e$, $\mu_{\nu_e}$ and $\{\mu_{fc}\}$.

%-------------------------------------------------------------------------
\section{Deconfinement transition in proto-neutron star matter}
%-------------------------------------------------------------------------

In order to determine the transition conditions, we apply the
Gibbs criteria, i.e. we assume that deconfinement will occur when
the pressure and Gibbs energy per baryon are the same for both
hadronic matter and quark matter at a given common temperature.
Thus, we have
\begin{eqnarray}
g^H = g^Q
\; , \qquad
P^H = P^Q
\; , \qquad
T^H = T^Q \;  ,
\label{gibbs}
\end{eqnarray}
where the index $H$ refers to hadron matter and the index $Q$ to quark matter. According
to these conditions (together with the equations of Sections 2 and 3), for a given temperature $T^H$ and neutrino chemical potential of the trapped
neutrinos in the hadronic phase $\mu_{\nu_e}^H$, there is an unique pressure $P$ at which the deconfinement is possible.
Instead of $P$, we may characterize the transition point by giving the Gibbs free energy per baryon $g$,
or alternatively, the mass-energy density of the hadronic phase $\rho_H$ (see Figs. 1-3).
We emphasize that, according to the present description, $P$ and $g$ are the same in both the hadronic phase
and the just deconfined phase. However, the mass-energy density $\rho_H$ and $\rho_Q$ at the transition point are different in general.
Similarly, while the abundance $Y_{\nu_e}$ of neutrinos is the same in both the hadronic and just  deconfined quark phases,
the chemical potentials $\mu_{\nu_e}^Q$ and $\mu_{\nu_e}^H$ are different.

According to numerical simulations \cite{Burrows1986,Keil1995,Pons1999},
during the first tens of seconds of evolution the protoneutron star cools from $T \sim 40$ MeV to temperatures
below 2-4 MeV. In the same period, the chemical potential $\mu_{\nu_e}^H$ of the trapped neutrinos
evolves from $\sim 200$ MeV to essentially zero.
Thus, in order to consider typical PNS conditions we have solved Eqs. (\ref{gibbs}) together with the equations of Sections 2 and 3
for temperatures in the range $0-60$ MeV and $\mu_{\nu_e}^H$ in the range $0-200$ MeV. The results
are displayed in Figs. 1-3 for all the parameterizations of the equations of state given in previous sections.

%++++++++++++++++++++++++++++
\begin{figure*}[t!]
\includegraphics[scale=.3]{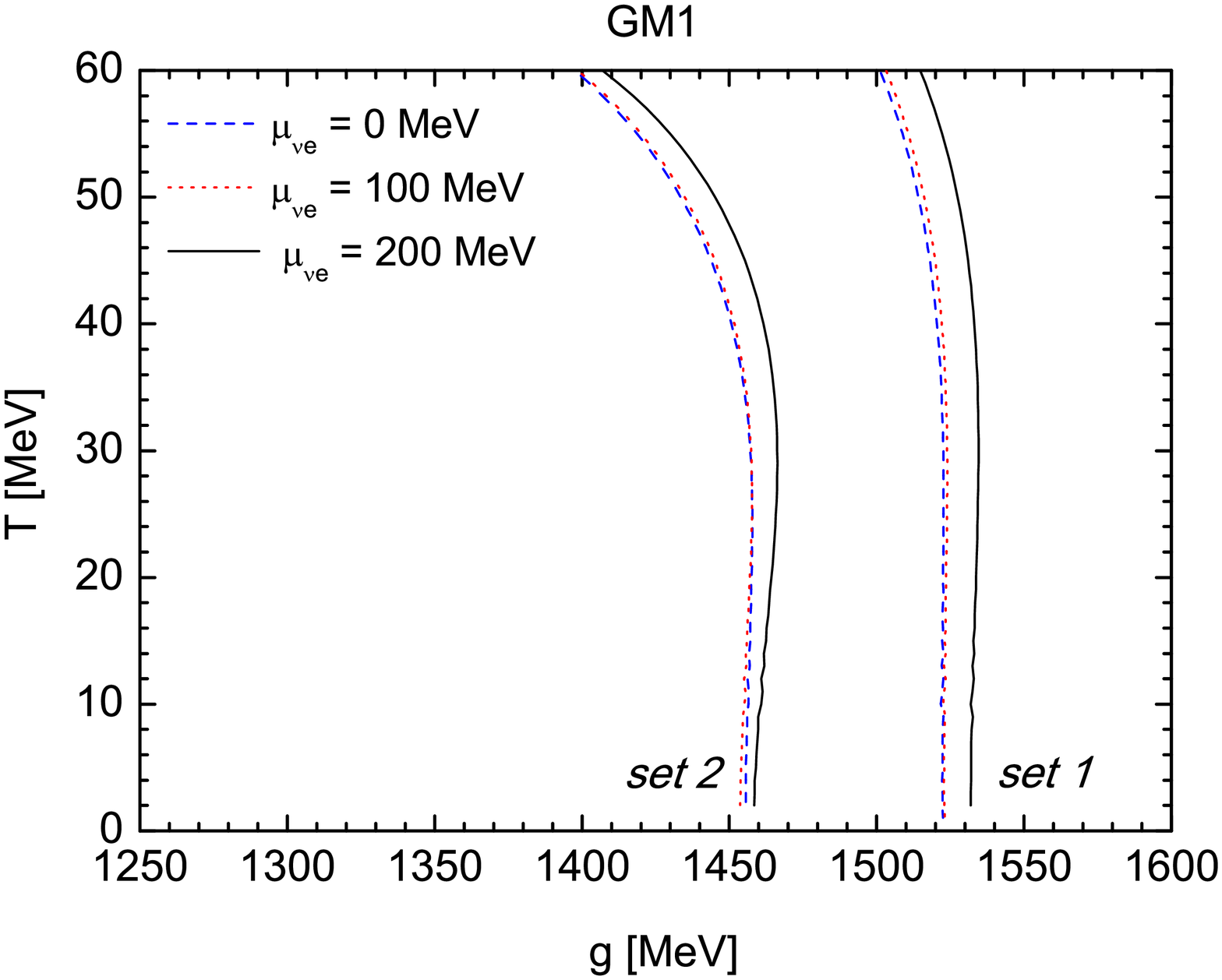}
\includegraphics[scale=.3]{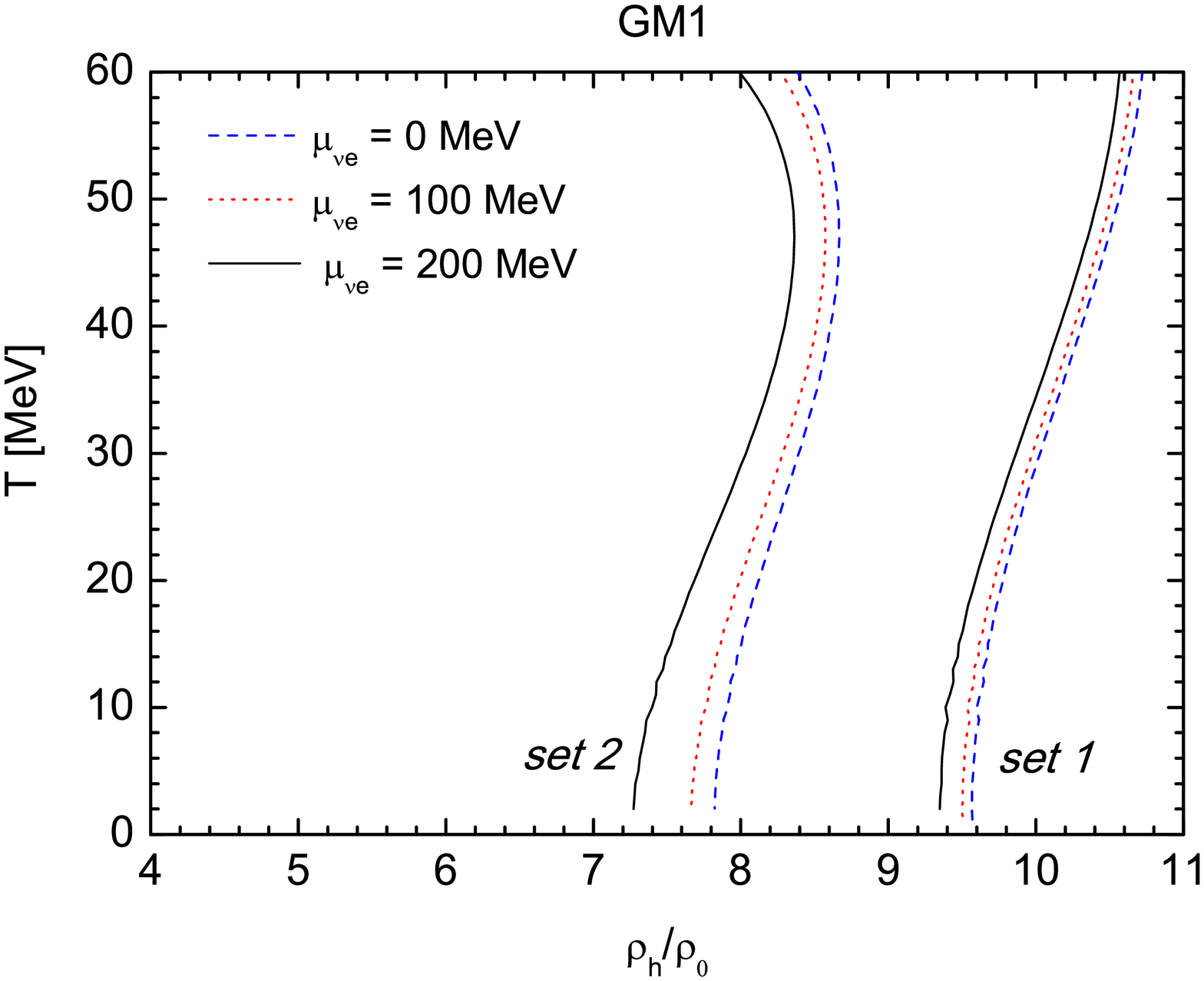}
\caption{\textit{Left panel:} the Gibbs free energy density per
baryon $g$ at which deconfinement occurs versus the
temperature $T$ for three different values of the neutrino chemical
potential in the hadronic phase ($\mu_{\nu_e}^H  = 0$ MeV in
dashed line, $\mu_{\nu_e}^H  = 100$ MeV in dotted line and
$\mu_{\nu_e}^H  = 200$ MeV in full line). \textit{Right panel:}
the mass-energy density of the  hadronic phase at which
deconfinement occurs versus the temperature $T$, for the same
values of $\mu_{\nu_e}^H$ given in the left panel (density is
given in units of the nuclear saturation density $\rho_0$ ). The
hadronic phase is described by the GM1 parametrization of the EOS.
For the quark phase we adopt the two parameterizations of the NJL
model given in Table II.  In both figures, if the thermodynamic state
of hadronic matter (characterized by \{ $T^H$, $g^H$, $\mu_{\nu_e}^H$\}
or by \{$T^H$, $\rho_H$, $\mu_{\nu_e}^H$ \} ) lies to the left of the curve
corresponding to the same $\mu_{\nu_e}^H$, then the deconfinement transition is not possible.
In the right side region of a given curve the preferred phase is deconfined quark matter.    } \label{fig1}
\end{figure*}
\begin{figure*}
\includegraphics[scale=.3]{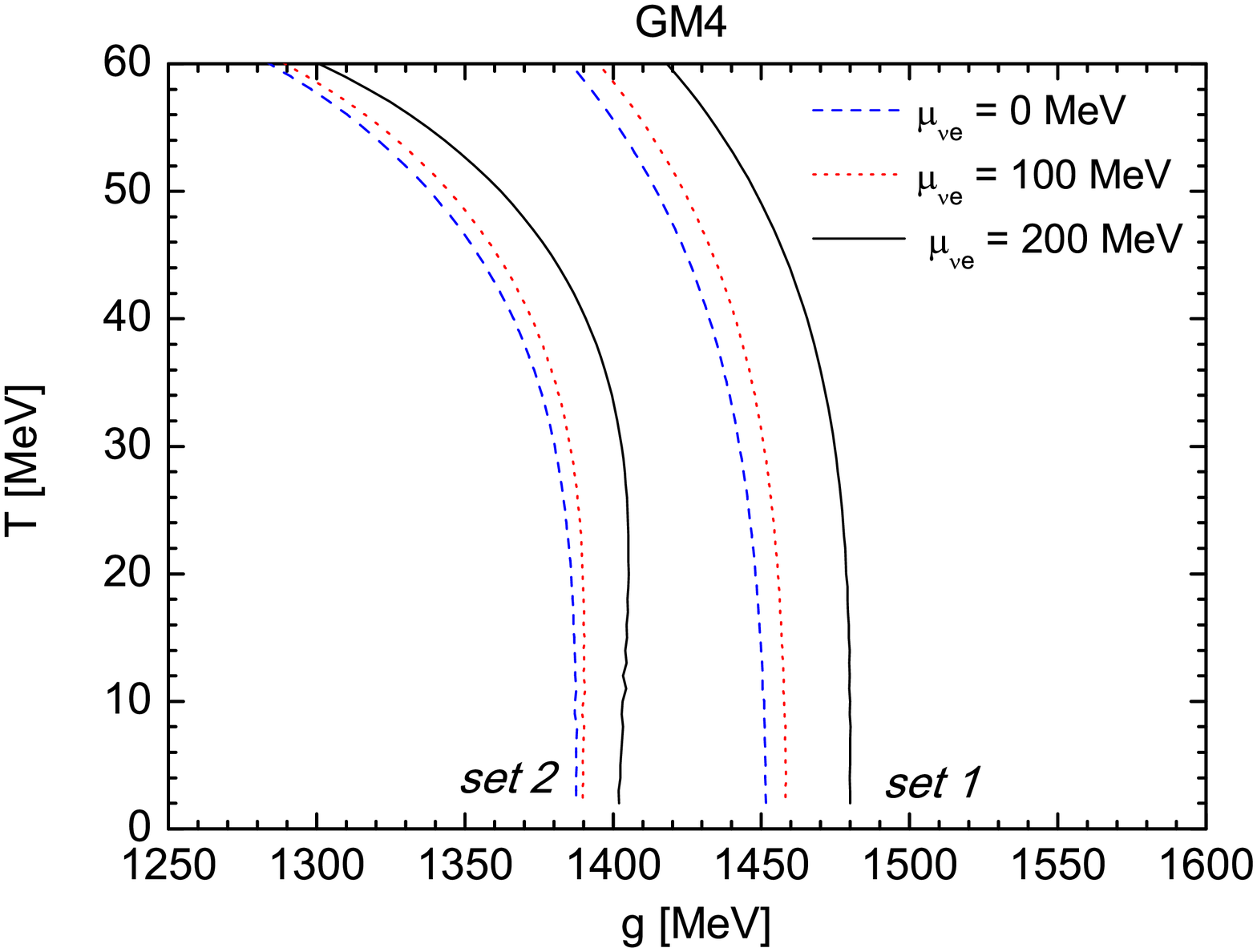}
\includegraphics[scale=.3]{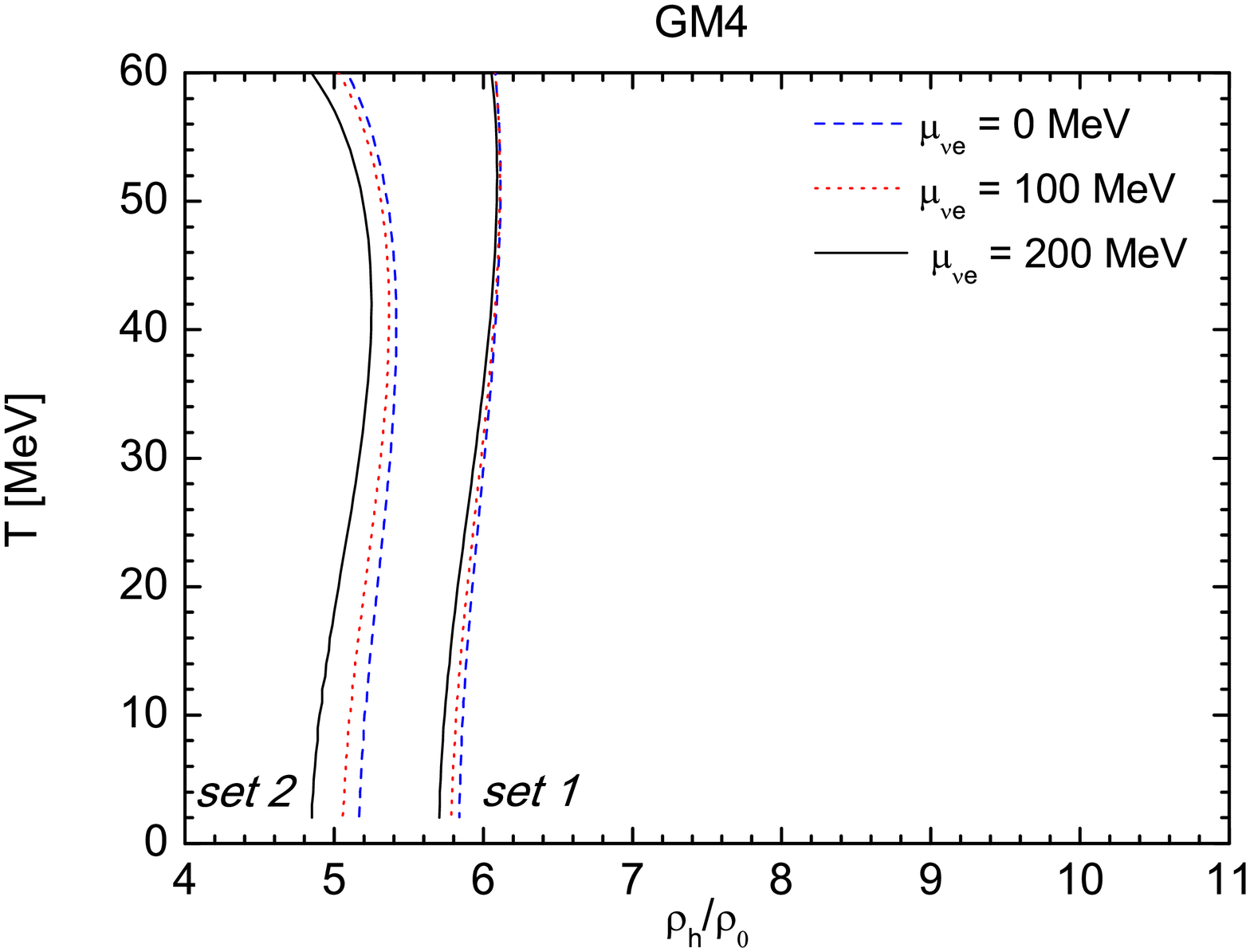}
\caption{Same as Fig. 1 but employing the GM4 parametrization of
the hadronic equation of state. } \label{fig2}
\end{figure*}

In Fig. \ref{fig1} we display the results for the GM1 parametrization of the hadronic EOS.
In the left panel of Fig. \ref{fig1} we show the Gibbs free energy density per
baryon $g$ at which deconfinement occurs versus the
temperature $T$ for three different values of the neutrino chemical
potential in the hadronic phase ($\mu_{\nu_e}^H  = 0,100,200$ MeV). In the right  panel the same results
are shown but as a function of the mass-energy density of the  hadronic phase (in units of the nuclear saturation density $\rho_0 =  2.7 \times 10^{14}$ g cm$^3$). In both figures, if the thermodynamic state  of hadronic matter (characterized e.g. by $T^H$, $\rho_H$ and  $\mu_{\nu_e}^H$) lies to the left of the curve corresponding to the same $\mu_{\nu_e}^H$, then the deconfinement transition is not possible. In the right side region of
a given curve the preferred phase is deconfined quark matter. Notice that the transition's Gibbs free energy is an increasing function of $\mu_{\nu_e}^H$. However, the transition density of the hadronic phase slightly decreases as $\mu_{\nu_e}^H$ increases.
In Fig. \ref{fig2} we display the results for the GM4 parametrization of the hadronic EOS.
The results are qualitatively the same but the transition densities are smaller than those for GM1 by $\sim 30 \%$.

In Fig. 3 we show the behavior of the transition's density as a
function of the chemical potential of trapped neutrinos
$\mu_{\nu_e}^H$ for two specific temperatures ($T=$  2 and 30
MeV). It is clearly seen that for a fixed temperature the effect
of deleptonization is to inhibit the transition. This effect is
not very large; at fixed temperature there is a slight increase by
less than a 10\% when $\mu_{\nu_e}^H$ falls from 200 MeV to 0 MeV.
On the other hand, the effect of cooling works in the opposite
direction because pairing tends to help the transition and the gap
increases as the temperature goes down. The effect of cooling is
also small; at fixed  $\mu_{\nu_e}^H$ there is a slight decrease
of the transition density $\rho_H$ by less than a 10\% when the
temperature falls from 30 MeV to 2 MeV. Both effects tend to
cancel each other as the PNS cools and deleptonizes, resulting a
transition density that is roughly constant along the evolution of
the protoneutron star.

\begin{figure*}[t!]
\begin{center}
\includegraphics[scale=0.3]{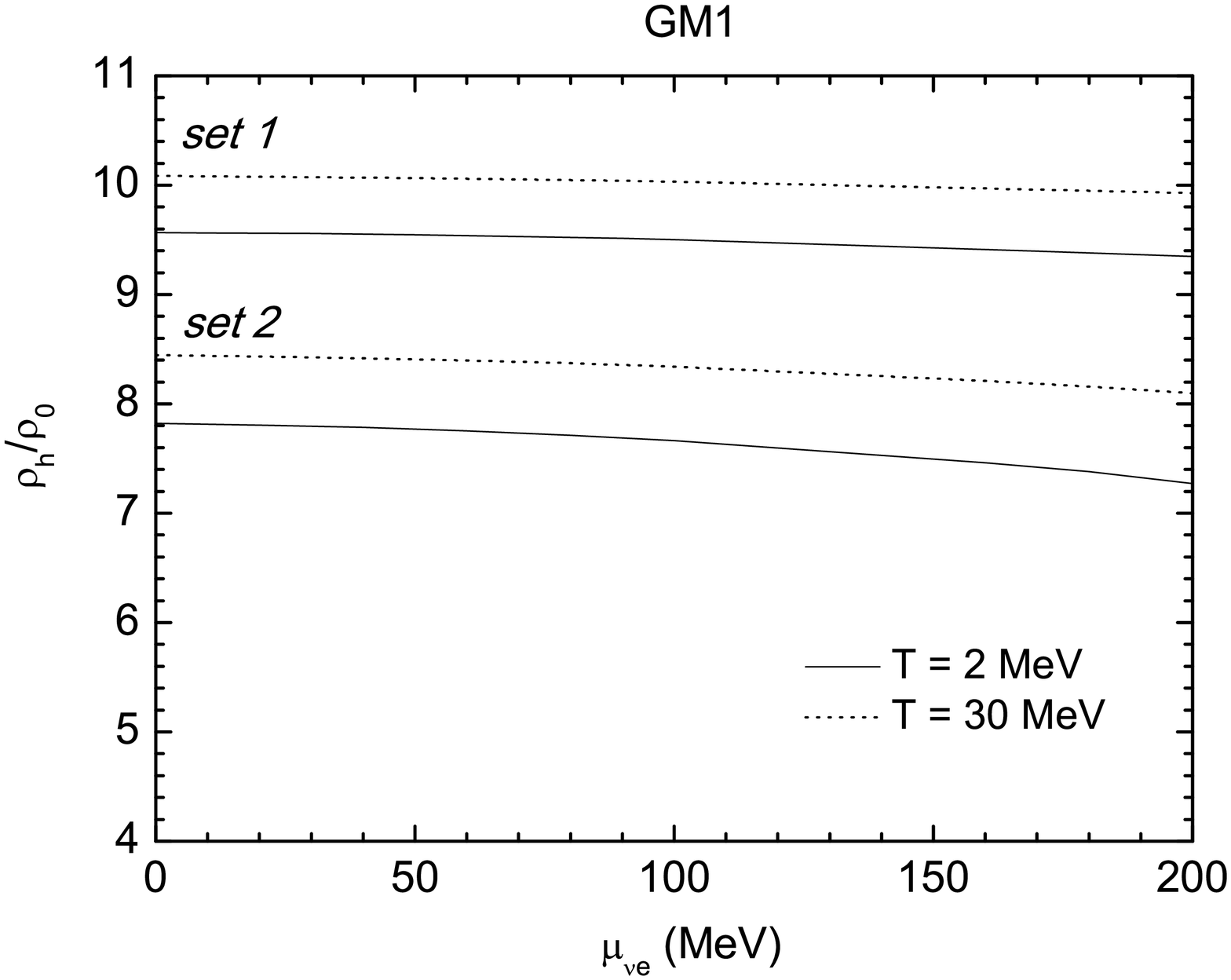}
\includegraphics[scale=0.3]{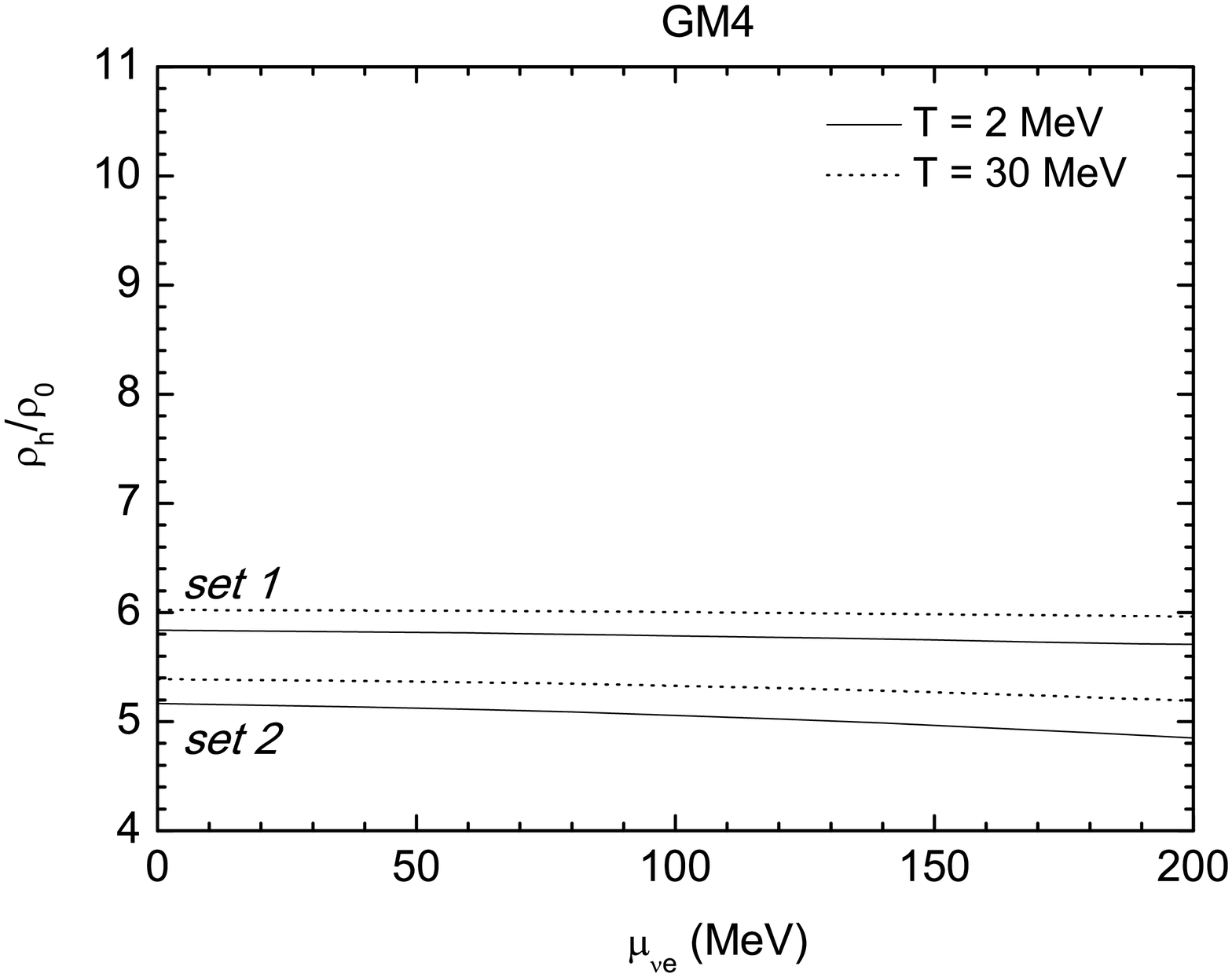}
\caption{The mass-energy density of the  hadronic phase at which
deconfinement occurs as a function of the chemical potential of
trapped neutrinos $\mu_{\nu_e}^H$. Results are given for two
temperatures: $T= 2$ MeV in full line and $T=30$ MeV in dashed
line. We employed the GM1 (left panel) and the GM4 (right panel)
parametrization of the hadronic EOS. Notice that there is a small
decrease of the transition density $\rho_{h}$ for large
$\mu_{\nu_e}^H$.}
\end{center}
\label{fig3}
\end{figure*}

%-------------------------------------------------------------------------
\section{Conclusions}
%-------------------------------------------------------------------------

In this paper we have investigated the role of color
superconductivity in the deconfinement transition from hadronic
matter to quark matter at finite temperature and in the presence
of a trapped neutrino gas. The study presented here is
relevant for  the first tens of seconds of evolution of newly born
protoneutron stars.

In our analysis we used a two phase description where we  employed the Nambu-Jona-Lasinio
model in the description of quark matter (Sec. III) and a non-linear Walecka model
which includes the whole baryon octet, electrons and electron neutrinos in equilibrium under weak
interactions in the description of hadronic matter (Sec. II).
Deconfinement is assumed to be a first order  phase transition and the just deconfined quark phase is assumed to
have the same ``flavor composition'' than the $\beta$-stable hadronic
phase from which it has been originated (see \cite{Lugones2009} and references therein). When color superconductivity is included
together with flavor conservation \cite{Lugones2005}, the most likely configuration of the just deconfined phase is 2SC provided the
pairing gap is large enough. This just deconfined phase is out of chemical equilibrium
under weak interactions and thus it is very short lived but it is a a kind of ``activation barrier'' that
determines the onset of the deconfinement transition.

The main result of the present paper is that, within the NJL model, deconfinement is more difficult
for small neutrino content and it is easier for lower temperatures.
This effect is not very large, at least for the here-used parameterizations of the NJL model.
At fixed temperature there is a slight increase by less than a 10\% when $\mu_{\nu_e}^H$ falls
from 200 MeV to 0 MeV (see Fig. 3).  The effect of cooling is also small; at fixed  $\mu_{\nu_e}^H$
there is a slight decrease of the transition density $\rho_H$ by less than a 10\% when the temperature
falls from 30 MeV to 2 MeV (see Fig. 3). This is due to the fact that the pairing gap becomes larger
as the temperature decreases and therefore the increase of the condensation term favors the transition
at low temperatures. Both effects tend to cancel each other as the PNS cools and deleptonizes,
resulting a transition density that is roughly constant along the evolution of the protoneutron star.

The here-found behavior is qualitatively opposite to what it was found within the MIT bag model. In fact,
previous analysis without including the effect of color superconductivity \cite{Lugones1998,Lugones1999}
show that the presence trapped neutrinos pushes up the transition density to values much larger than for
neutrino free matter. It was also found in \cite{Lugones1998,Lugones1999} that the transition is easier
for larger temperatures. More recent results including the effect of color superconductivity
within the MIT Bag model \cite{Taiza2010} show that the transition density increases with neutrino trapping
but (in coincidence with the here found results) the pairing gap favors the transition as the temperature decreases.

In spite of some differences between the results within the NJL and the MIT bag model description of quark matter
some general conclusions may be obtained about the effect of color superconductivity in the deconfinement transition.
First, when color superconductivity is present the deconfinement density is not so strongly affected
by neutrino trapping as it is in the unpaired case. Second, color superconductivity makes the transition easier
at lower temperatures and the dependence of the deconfinement density with $T$ is much smaller than
in the unpaired case.

During cooling and deleptonization of the protoneutron star the temperature and
the chemical potential of trapped neutrinos fall abruptly in a few seconds and
there is also some contraction of the whole neutron star. It is interesting to
note that although the density increase is not too large, it may be comparatively
important for the deconfinement transition because the effects of temperature and
neutrino trapping are smoothed by color superconductivity. According to our
results the deconfinement transition is favored after substantial cooling and
contraction of the protoneutron star but full numerical simulations of
protoneutron star evolution are needed in order to determine whether and when the
deconfinement conditions are attained.

\section{Acknowledgements}

This work was supported in part by CONICET (Argentina) grant \#
PIP 6084 and by ANPCyT  (Argentina) grant \# PICT07 03-00818. T.
A. S. do Carmo acknowledges the financial support received from
UFABC (Brazil). G. Lugones acknowledges the financial support
received from FAPESP (Brazil).


\begin{thebibliography}{99}

\bibitem{Burrows1986}  A. Burrows and J. M. Lattimer,  Astrophys. J. {\bf 307}, 178 (1986).

\bibitem{Keil1995} W. Keil and H-Th. Janka,  Astron. Astrophys., 296, 145 (1995).

\bibitem{Pons1999} J. A. Pons et al., Astrophys. J.  513, 780, (1999).

\bibitem{Lugones1998} G. Lugones and O. G. Benvenuto, Phys. Rev. D {\bf 58}, 083001 (1998).

\bibitem{Lugones1999} O. G. Benvenuto  and  G. Lugones,  Mon. Not. R.A.S.  {\bf 304}, L25 (1999).

\bibitem{IidaSato1998} K. Iida and K. Sato, Phys. Rev. C {\bf 58}, 2538 (1998).

\bibitem{Madsen1994} M. L. Olesen and J. Madsen, Phys. Rev. D {\bf 49}, 2698 (1994).

\bibitem{Bombaci2004}  I. Bombaci, I. Parenti, I. Vida\~na, Astrophys.J. {\bf 614}, 314 (2004).

\bibitem{Lugones2005} G. Lugones and I. Bombaci, Phys. Rev. D {\bf 72}, 065021 (2005).

\bibitem{Bombaci2007}  I. Bombaci, G. Lugones, I. Vida\~na, Astronomy and Astrophysics {\bf 462}, 1017 (2007).

\bibitem{Bombaci2009} I. Bombaci, D. Logoteta, P.K. Panda, C. Providencia, I. Vidana,  Phys. Lett. B \textbf{680}, 448 (2009)

\bibitem{Lugones2009} G. Lugones, A. G. Grunfeld, N.N. Scoccola and C. Villavicencio, Phys. Rev. D \textbf{80}, 045017 (2009)

\bibitem{GM1} N. K. Glendenning and S.A. Moszkowski, Phys. Rev. Lett. {\bf 67}, 2414 (1991).

\bibitem{qhd} J.D. Walecka, Ann. Phys. {\bf 83}, 491 (1974);  B.D. Serot and J.D. Walecka, Adv. Nucl. Phys. {\bf 16}, 1 (1986).

\bibitem{Menezes} D.P. Menezes and C. Provid\^encia, Phys. Rev. C {\bf 68}, 035804 (2003); A.M.S. Santos and D.P. Menezes, Phys. Rev. C {\bf 69}, 045803 (2004).

\bibitem{mfm} N. K. Glendenning, Astrophys. J. {\bf 293}, 470 (1985).

\bibitem{Huang:2002zd} M.~Huang, P.~f.~Zhuang and W.~q.~Chao, Phys.\ Rev.\  D {\bf 67}, 065015 (2003).

\bibitem{Ruester:2005jc} S.~B.~Ruester, V.~Werth, M.~Buballa, I.~A.~Shovkovy and D.~H.~Rischke, Phys.\ Rev.\  D {\bf 72}, 034004 (2005).

\bibitem{Blaschke:2005uj} D.~Blaschke, S.~Fredriksson, H.~Grigorian, A.~M.~Oztas and F.~Sandin, Phys.\ Rev.\  D {\bf 72}, 065020 (2005).

\bibitem{Rehberg:1995kh} P.~Rehberg, S.~P.~Klevansky and J.~Hufner, Phys.\ Rev.\  C {\bf 53}, 410 (1996).

\bibitem{Hatsuda:1994pi}  T.~Hatsuda and T.~Kunihiro, Phys.\ Rept.\  {\bf 247}, 221 (1994).

\bibitem{Buballa2005} M. Buballa, Phys. Rept. {\bf 407}, 205 (2005).

\bibitem{Taiza2010} T. A. S. do Carmo and G. Lugones, to be submitted.





\end{thebibliography}
\end{document}